\documentclass[trackchanges]{aastex701}
\usepackage{subfigure}
\usepackage{amsmath}
\usepackage{caption}

\begin{document}

\title{A possible mechanism to explain the prograde equatorial jet of a Jupiter-like gaseous giant}

\author[orcid=0009-0000-4927-8976,sname=Lian]{Yuchen Lian}
\affiliation{Shanghai Astronomical Observatory, Chinese Academy of Science, No 80. Nandan Road, Shanghai, China, 200030}
\email{lianyuchen@shao.ac.cn}  

\author[orcid=0000-0002-1365-0577,gname=Pengshuo, sname=Duan]{Pengshuo Duan} 
\affiliation{Shanghai Astronomical Observatory, Chinese Academy of Science, No 80. Nandan Road, Shanghai, China, 200030}
\email{duanps@shao.ac.cn}

\author[orcid=0000-0003-2339-0294,gname=Dali,sname=Kong]{Dali Kong}
\affiliation{Shanghai Astronomical Observatory, Chinese Academy of Science, No 80. Nandan Road, Shanghai, China, 200030}
\email[show]{dkong@shao.ac.cn}

%% Use the \collaboration command to identify collaborations. This command
%% takes an optional argument that is either a number or the word "all"
%% which tells the compiler how many of the authors above the command to
%% show. For example "\collaboration[all]{(DELVE Collaboration)}" wil include
%% all the authors above this command.
%%
%% Mark off the abstract in the ``abstract'' environment. 
\begin{abstract}
Gaseous giants are characterized by their deep atmospheres, which lack clear boundaries with their interiors; therefore, their internal states could directly influence atmospheric dynamics. So far, most modeling studies have considered deep convection as the primary mechanism by which the interior influences atmospheric dynamics. In this work, we propose another possible mechanism that might crucially determine the appearance of gaseous giants’ atmospheric cloud-top jet winds, tracing them to a typical hydromagnetic wave (the so-called equatorial Magnetic-Archimedes-Coriolis wave) generated within the stably stratified, strongly magnetized helium rain layer. The associated thermal perturbations can propagate upward through the convective molecular hydrogen envelope, eventually affecting the atmospheric thermal structure—the zonal inhomogeneities that are conducive to the formation of the eastward atmospheric equatorial jet (super-rotation). Our results have important implications for understanding the equatorial dynamics of gaseous giants. This mechanism could also help explain the equatorial westward jets (sub-rotation) observed on Uranus and Neptune, {which lack the helium rain layers}.
\end{abstract}


\section{Introduction} 
The atmosphere of gas giants can smoothly transition into the interior. On a large scale, as pressure increases from the surface to the core, theories predict that gaseous giants are composed of the following layers: the weather layer, molecular hydrogen layer (MoHL), metallic hydrogen layer, and core, without rigid boundaries between them \citep{mil-2016,miguel,ni}. Still, a substantial dynamic gulf exists between the interior and the atmosphere. {The atmospheric winds, namely a strong and wide equatorial eastward (prograde) jet together with multiple off-equatorial belts and zones \citep{he-2003}, such as Jupiter \citep{limaye,porco,garcia,tollefson2017}, and Saturn \citep{smith,san-2000,li,garcia-2011}}, exhibit distinct spatial inhomogeneities, {while the interiors of gaseous giants are assumed to be isentropic and horizontally homogeneously mixed \citep{helled-2018,miguel}}. Hence, the influence of the interior dynamics of gaseous giants on their atmospheres remains a significant research question. 

{A number of studies \citet{zhang-b,kaspi-etal-2009,christensen-2020,gastine-wicht-2021,heimpel,duer-2023,wulff-2022,wulff,christensen-2024,currie} explored the deep convective influence on the atmosphere dynamics with barotropic conditions that the jets obey the Taylor-Proudman constraint, arranged by fluid columns which are parallel to the axis of rotation.} {The columns stretching lead to eastward-propagating Rossby waves, which carry eastward momentum and maintain equatorial eastward jets \citep{lemasquerier}.} {Nevertheless, this scenario does not explain the dynamics in the weather layer, located above the radiative-convective boundary,} for example, the quasi-quadrennial oscillations of Jupiter \citep{leovy}.  Furthermore, the location and number of off-equatorial jets are directly related to the depth of the fluid spherical shell containing the Taylor-Proudman column \citep{duer-2023}. The midlatitude jets only appear when the depth of the simulation is limited. Some simulations need additional lower boundary conditions to achieve multiple midlatitude jets \citep{yadav}.

Some studies suggested that the interior dynamics' impact on the atmosphere is not significant, and that the atmospheric jets are confined within the shallow atmospheric weather layer \citep[e.g.][]{mcintyre2016}. The driving forces of the jets arise from the temperature gradients and baroclinic instabilities that lead to Reynolds stress \citep{lian2008,s11,spiga}, or from sporadic thunderstorms \citep{showman2006}, etc. The number of off-equatorial jets is easily captured in General Circulation Model (GCM) simulations. However, to achieve an eastward (prograde) equatorial jet, additional heat sources must be required \citep{willams,lian2008}, {and the wind strength of these jets decays with depth in a much shallower region, which contradicts the measurements of Jupiter from spacecrafts \citep{kaspi-etal-2018}.} 

Beyond the two scenarios described above, inertial wave modes within the convective zone can propagate to the surface \citep[e.g.][]{blume}, potentially influencing the weather layer’s dynamics. Here, we propose a new coupling mechanism for jet origin, by considering the effects of a type of hydromagnetic waves, called the equatorial Magnetic-Archimedes-Coriolis (eMAC) waves, which are just generated in a stable helium rain layer (HRL) of gaseous giants' deep interiors, and the wave perturbation propagates upward through the MoHL to the weather layer. In this explanation, perturbations in eMAC waves can induce an equatorial zonal inhomogeneity, which, in turn, triggers equatorial prograde jets in the atmosphere. Similar to the Matsuno-Gill mode \citep{mat-1966,gill}, this mechanism requires zonal inhomogeneities for equatorward momentum transport to drive prograde jets. In synchronous rotating hot Jupiters, day-night temperature contrast generates zonal inhomogeneities, leading to the common occurrence of prograde jets \citep{s15,plur,lesjak}. By contrast, in the gaseous giants discussed in this paper, such inhomogeneities are generated by the perturbation of hydromagnetic waves.

The paper is organized as follows. We lay out the theory in Section \ref{sec:theory}, including the eMAC waves (Section \ref{sec:emac}) and waves in MoHL (Section \ref{sec:MoHL}). We introduce the numerical model in Section \ref{sec:model}. In Section \ref{sec:result}, we present the results and discuss the associated mechanisms. Discussion is in Section \ref{sec:dis}. 

\section{Theory}
\label{sec:theory}
\subsection{The eMAC waves in the helium rain layer}
\label{sec:emac}
After the formation of gaseous giants, their interiors undergo a long cooling process accompanied by a homogeneous to non-homogeneous transition: helium precipitates from hydrogen, resulting in the formation of HRL \citep{mor-2013,mil-2016,brygoo}. {The ``helium rain'' falls into the interior, leaving latent heat and a compositional gradient at the precipitation region, causing double-diffusive motions \citep[e.g.][]{leconte-chabrier-2012,nettelmann-2015,fuentes-2022}, where the latent heat leads to a convectively unstable superadiabatic temperature profile, while the compositional gradient tends to counteract the superadiabatic convective instability \citep{mankov}. As a result, in principle, some part of an HRL could be stably stratified.} 

{Recent dedicated research does support the existence of stable stratification within double-diffusive HRL, for example, (1) The temperature gradients of the HRL in Jupiter and Saturn do not exceed the maximum super-adiabatic temperature gradient that is permitted by stable stratification \citep{markham}. This phenomenon can be attributed to the latent heat carried by helium raindrops traversing the stable layer, thereby reducing the temperature gradient \citep{leconte-2024}. (2) The helium raindrop settling speed could be larger than the buoyant speed \citep{mankov,friedson-gonzales}, thereby suppressing convection. (3) Within the HRL, the effective coefficient of thermal expansion is negative, which further inhibits convective processes  \citep{markham}. In addition, the constraints of gravity fields, dynamo models, and seismology favor stable stratifications in Jupiter \citep{moore2018,tsang-jones} and Saturn \citep{mankovich-fuller}.} 

%{Now that a stable layer is formed within HRL of gaseous giants \citep{ledoux,leconte-2017,vazan}, and the} 
The HRL of gaseous giants is widely accepted as being situated in proximity to the outer boundary of the metallic hydrogen layer \citep{zaghoo}. The stable layers within HRL resemble a ``sandwich'' structure, lying within the above and below convective layers, and enable the eMAC waves \citep{buff-2019,chi-2020,duan-2023}. Note that the eMAC waves are only confined within the stable layer and they are produced by the interplay between magnetic, buoyancy and Coriolis forcings. {The eMAC waves may also form within the double-diffusive layers, provided that local convective stability is permitted}. These hydromagnetic waves are represented by the perturbed magnetic field signals governed by the Weber equation, here the two-dimensional spatial structure of the perturbed magnetic field \citep{duan-2023} is described by:

\begin{equation}
\tilde{B}_{\theta}\propto e^{-1/2\tilde{\alpha}^2\hat{y}^2}H_n(\tilde{\alpha}\hat{y})
\label{eqn:m1}
\end{equation}
where $\tilde{B}_{\theta}$ is normalied magnetic field perturbations at meridional direction, $\hat{y}=\cos\theta$ is a meridional coordinate with colatitude $\theta$, $\tilde{\alpha}=\alpha_2^{1/4}$ and $H_n$ is Hermite polynomial solution. $\alpha_2$ is related to zonal wavenunber $m$, stable-layer thickness $H$, buoyancy frequency $N_1$ and meridional gradient factor of magnetic field $\Lambda$. For the value of parameters, please refer to Section \ref{sec:emacpara}. For the full derivation of eMAC waves, please refer to the Appendix \ref{sec:A_emac}.

The eMAC wave could generate zonal inhomogeneity \citep{buff-2019,duan-2023}. However, given that the eMAC wave only exists in HRL, while the MoHL lies between HRL and the weather layer, it is required to answer the question of whether the perturbation of eMAC waves could propagate through MoHL. 

\subsection{Upward propagation of the eMAC wave perturbation}
\label{sec:MoHL}

The perturbations induced by the eMAC waves can propagate through MoHL up to the bottom of the weather layer, despite the absence of eMAC waves in the fully convective MoHL above HRL \citep{mil-2016}. This layered propagation physics bears similarity to the radial propagation characteristics of inertia-gravity waves and thermal Rossby waves in stellar tropospheres \citep{roberts-1968,busse-1970}. Previous studies show some kinds of inertial oscillations within a star \citep{papal}. In a stellar rotating spherical shell, a spinning column near the equator grows in height when it moves toward the rotation axis, and it will spin faster to conserve angular momentum. The new flow field causes the west side column to move away from the rotation axis and the east side column inward, the process of which exhibits the prograde Thermal Rossby wave propagation \citep{busse-1970}. Such wave modes, called $r$-modes, whose frequencies are selected by the stellar convective zone \citep{cai-2021,bekki-etal-2022,jain-2023,albekioni-2023}. Therefore, planetary convective MoHL could also display selective behavior for wave modes with frequency $\omega$.

According to the previous studies \citep[e.g.][]{hin-2022}, we solve the perturbation in MoHL as a plane-wave form of $e^{{\rm i}(m\lambda-\omega t)}$, where $\lambda$ is longitude, $\omega$ is frequency, and $t$ is time. Then we give a threshold frequency $\omega_c$ (see Appendix \ref{sec:A_MoHL}):

\begin{equation}
    \omega_c\equiv\frac{2\Omega m}{[(\frac{\omega_{ac}}{c_s})^2+(\frac{m}{a})^2]aH_{\rho}}
    \label{eqn:m3}
\end{equation}
where $\omega_{ac}=\frac{c_s}{2H_{\rho}}\sqrt{1-2\frac{dH_{\rho}}{dz}}$ is the acoustic cutoff frequency, $\Omega$ is the planetary rotation velocity, and $a$ is the equatorial radius. $z$ is the vertical coordinate that $z=s-a$, and $s$ is the radius coordinate in cylindrical coordinates. The sound speed $c_s$, and the density scale height $H_{\rho}$ in MoHL are calculated by the background density $\rho_0$ and background pressure $p_0$. For the full derivations, please refer to the Appendix \ref{sec:A_MoHL}.

Eqn. \ref{eqn:m3} defines a waveguide constraint whereby waves with frequencies below the threshold frequency $\omega_c$ propagate transparently in the MoHL, while those above $\omega_c$ undergo dissipation. Therefore, in the event of an eMAC wave mode possessing an eigenfrequency lower than $\omega_c$, the spatial perturbations propagate through the MoHL to reach the bottom of the weather layer and influence the lower boundary conditions. Consequently, the atmospheric bottom boundary conditions in a general circulation model show the same spatial pattern as the eMAC wave mode at the upper boundary of HRL.

\section{Model}
\label{sec:model}
To simulate the impact of internal forcing on the atmospheric dynamics, we then employ a popular general circulation model, MIT GCM (MITgcm) \citep{adcroft}, which is an atmospheric and oceanic model that was developed by Massachusetts Institute of Technology. MITgcm calculates the atmospheric circulation by the governing global hydrostatic primitive equations in Appendix \ref{sec:A_dyn}. The model features a horizontal resolution of~$0.7^\circ$ and consists of 100 vertical levels spanning pressures from~$10^{-4}$ to 100~bars {(depth of $\sim$ 200 km to -250 km, let 1 bar corresponds to 0 km)}. The time step for the simulation is set to 100 seconds. 

We apply the Newtonian cooling scheme as the radiation scheme, where the difference between the local temperature and the radiation equilibrium temperature $T_{\rm eq}$ is utilized to calculate the heating rate. The temperature field relaxes to $T_{\rm eq}$ on the timescale of the relaxation time, $\tau_{\rm rad}$, which is the Newtonian cooling scheme. An Ohmic drag term $-\textbf{u}_{\rm h}/\tau_{\rm drag}$ was incorporated at the bottom, where $\textbf{u}_{\rm h}$ is horizontal velocity and $\tau_{\rm drag}$ is drag timescale.

The eMAC-wave-induced forcing is parameterized as a heating term $S_{\rm eMAC}$ in the thermodynamic equations, added at the bottom of the model. Besides, the latent heat released by the cloud is parameterized as another heating term $S_{\rm rand}$, which is an isotropic and random forcing added at the cloud level. 

\subsection{eMAC-wave-forcing parameters}
\label{sec:emacpara}
\begin{figure}[ht]
\captionsetup{labelfont={bf,color=red}}
	\centering
	\includegraphics[width=0.95\textwidth]{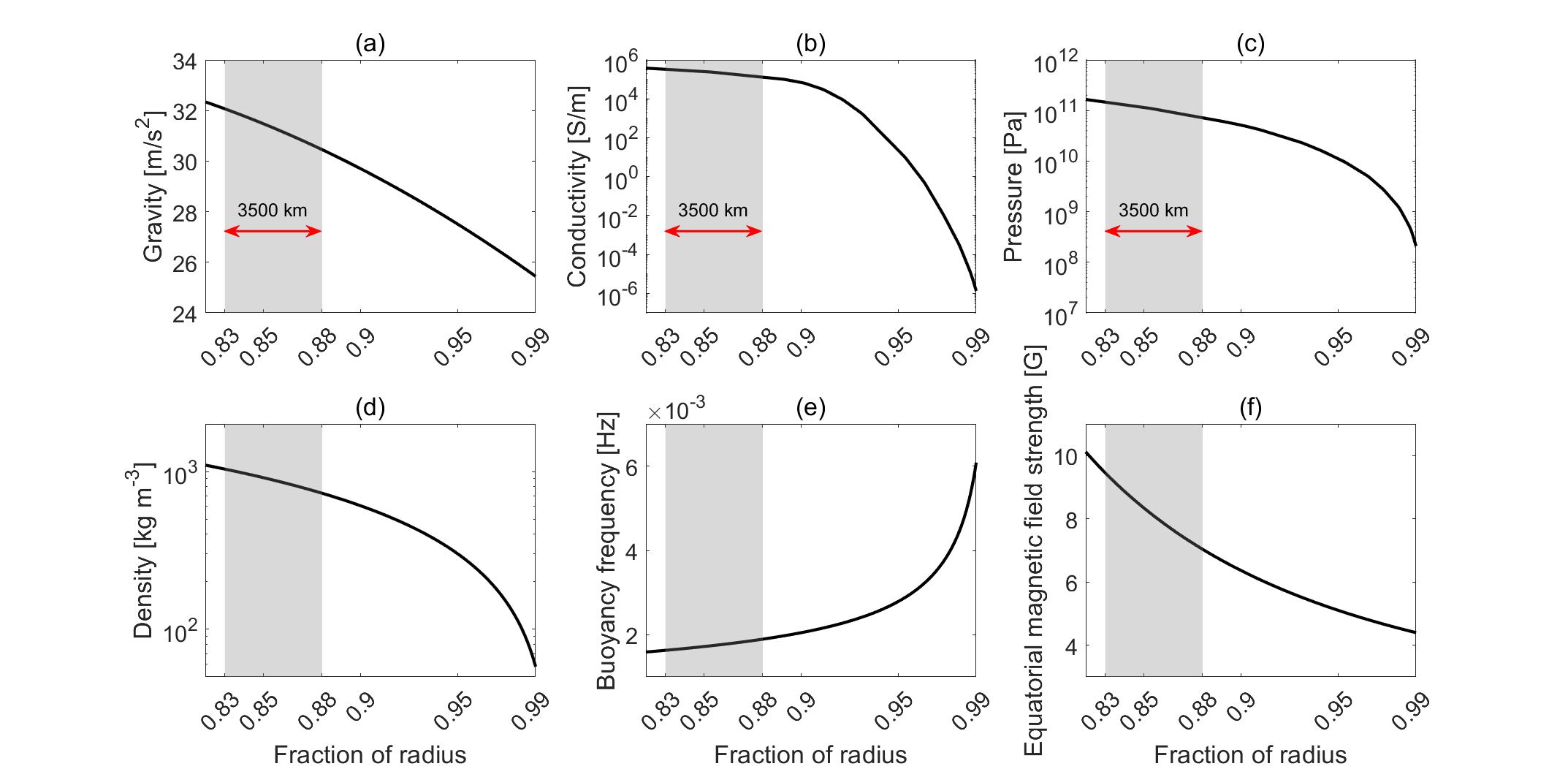} 
	\caption{The Jupiter-like gaseous giant reference states considered in our study. (a) Gravity $g$ \citep{gastine-wicht-2021}, (b) Electrical conductivity $\sigma$ \citep{french-etal-2012}, (c) Pressure $p_0$ \citep{mil-2016}, (d) Density $\rho_0$ \citep{helled-2018}, (e) Buoyancy frequency $N$, and (f) Equatorial magnetic field strength $B_0$ \citep{connerney,zaghoo} as a function of the fraction of radius. The shaded area is HRL.}
	\label{fig:e1}
\end{figure}

We adopted planetary parameters relevant for a Jupiter-like gaseous giant, which are provided in Figure \ref{fig:e1}, including gravity $g$, electrical conductivity $\sigma$, background pressure $p_0$, background density $\rho_0$, buoyancy frequency $N$, and equatorial magnetic field strength $B_0$. 

Estimating the thickness of HRL has been the focus of numerous studies; nevertheless, the exact thickness of the stable layer remains unclear. HRL has been predicted with an outer boundary 0.86 fraction of Jupiter's radius \citep{zaghoo,mankovich_fortney}, and within a range of 0.85 to 0.9 fraction of radius (equivalent to 3500 km thick) \citep{stevenson}, 0.78 to 0.86 fraction of radius (equivalent to 4000 km thick) \citep{hubbard-2016}, 0.87 to 0.91 fraction of radius (3000 km thick) \citep{mor-2013}, and 0.9 to 0.95 fraction of radius (3500 km thick) \citep{moore-etal-2022}. {The Juno probe estimated the dynamo radius is 0.81 \citep{connerney} or 0.807 \citep{wulff-2025} fraction of the radius. \citet{wulff-2025} imposed radical constraints on the stable stratification: the inner boundary cannot extend deeper than 0.8 fraction of radius, and the outer boundary cannot extend shallower than 0.9 fraction of radius.}

{The thickness of the stable layer may be thousands of km \citep{mil-2024},} or be much thinner, with a lower limit of thickness, about a hundred km \citep{markham}. The majority of studies have consensus that the outer boundary of HRL will not be shallower than 0.9 fraction of radius. In this study, we employ a value of 0.83 to 0.88 fraction of radius {($\sim$ 1.5 to 0.7 Mbar)} to predict the location of HRL, and we choose 3500, 2000, and 1000 km as the estimation of the stable-layer thickness $H$.

{The spherical harmonic internal magnetic field model JRM33 was utilised to estimate the radial magnetic field strength in the equatorial region (Figure \ref{fig:e1}(f)) \citep{stevenson1983,gastine-2014,duarte,gastine-wicht-2021,wilson}}. Jupiter's magnetic field strength at the poles is greater than at the equator \citep{connerney}, and we set $\Lambda=2$ to fit the latitudinal distribution.

\begin{figure}[ht]
    \centering  
    \subfigure[]{  
        \includegraphics[width=0.45\textwidth]{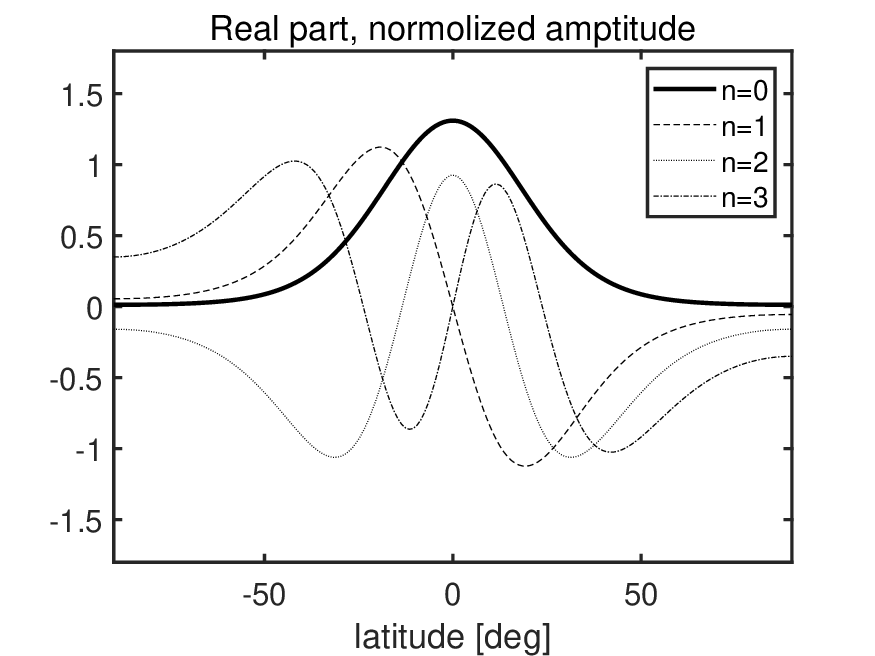}  
        \label{fig:11}  
    }  
    \hfill  
    \subfigure[]{  
        \includegraphics[width=0.45\textwidth]{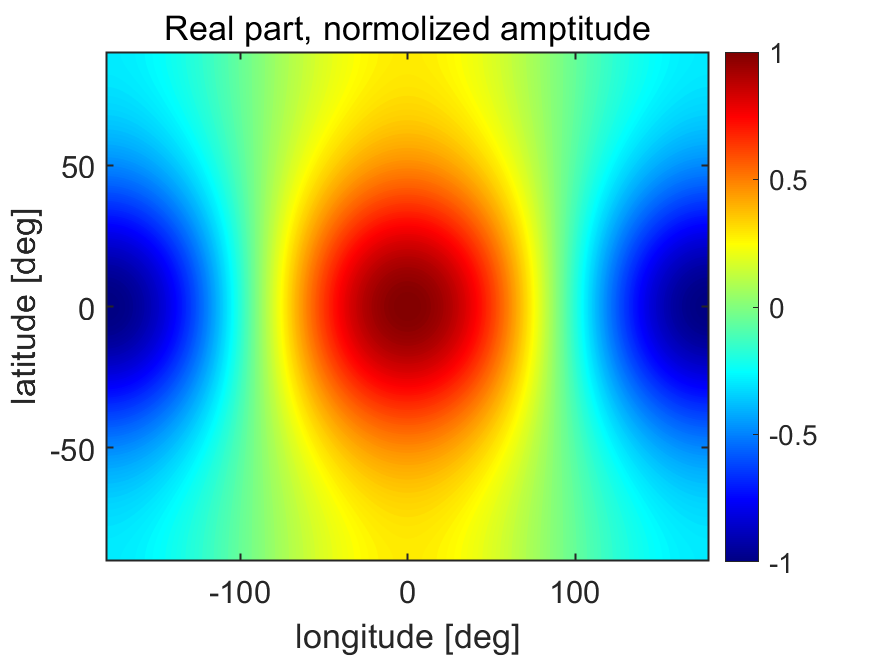}
        \label{fig:12}  
    }  
    \caption{Equatorial confinement of eMAC waves. (a) Latitudinal distribution of eMAC waves in Eqn.~\ref{eqn:m1} ($\tilde{\alpha}$ is set to be the real numbers). (b) Sketch of the normalized eMAC waves on the longitude-latitude plane with a stable-layer thickness of 3500 km and $m=1$.}  
    \label{fig:m2}  
\end{figure}

Figure \ref{fig:11} shows the analytical solution of Eqn. \ref{eqn:m1} with Hermite polynomial degree $n=0-3$. It is important to note that  Eqn. \ref{eqn:m1} is actually a truncation treatment that removes higher-order terms $O$($\hat{y}^4$), which performs well in the equatorial region, particularly within latitude~$\sim\pm 25^\circ$ \citep{duan-2023}. As latitude increases, the eMAC wave ceases to possess an analytical solution. In the case where $n > 0$, the energy of eMAC waves is more concentrated in high-latitude regions, which causes large relative errors of Eqn. \ref{eqn:m1}. We focus on the low-latitude dynamics, therefore, we consistently assign $n=0$ in the subsequent calculations. Figure \ref{fig:12} shows an example of an eMAC wave confined to latitude~${\pm 40^\circ}$, with $m=1$ and a thickness of stable layer $H=3500$~km. Crucially, we find that $n=0$ mode can give the ideal results (the following sections), suggesting that the specific eMAC wave solution (i.e., $n=0$) indeed exists to explain the observations, through the other modes with higher degrees may also exist.

The low-order analytical real solutions of eMAC waves have two eigenvalues of frequencies, calculated by Eqn. \ref{eqn:14}. These eigenfrequencies are about $\sim10^{-11}$ ${\rm s^{-1}}$ when $n=0$ is applied. In comparison, the threshold frequency in MoHL, denoted by $\omega_c$ in Eqn. \ref{eqn:m3}, is approximately $10^{-6}$ ${\rm s^{-1}}$ when $m = 1$, and approximately $10^{-5}$ ${\rm s^{-1}}$ when $m = 4$. It is evident that the threshold frequency increases with the zonal wavenumber, and all of the threshold frequencies are much higher than the eigenfrequency of eMAC waves in our simulations, meaning that the perturbations of eMAC waves are transparent in MoHL. Therefore, the eMAC wave-induced perturbations could propagate to the weather layer of a Jupiter-like gaseous giant.

\subsection{Atmospheric parameters}
We adopted atmospheric parameters for a Jupiter-like gaseous giant, including surface gravity $g_{\rm s}=24.79$ m ${\rm s^{-2}}$, specific heat $c_p=1.3 \times 10^4$ J ${\rm kg^{-1}}$ ${\rm K^{-1}}$ and specific gas constant $R_{\rm gas} = 3714$ J ${\rm kg^{-1}}$ ${\rm K^{-1}}$. In the Newtonian cooling radiative scheme, the equilibrium temperature $T_{\rm eq}$ and radiative timescale $\tau_{\rm rad}$ depend solely on the pressure coordinate, with specific values being determined by Jupiter's reference state delineated in \citet{li-etal-2018}. The radiative timescale at the bottom layer (100 bars) is very short (100 seconds) for the fixed thermal boundary conditions.

The eMAC-wave-induced forcing corresponds to $S_{\rm eMAC}$, which is a fixed heating from the bottom (100 bars) to 1 bar with the amplitude exponentially decreasing from~$10^{-7}$~${\rm K/s}$ to zero, with a spatial pattern as Figure \ref{fig:12}. Because the eigenfrequencies of the eMAC wave ($10^{-11}$ ${\rm s^{-1}}$) are much lower than the rotation rate ($10^{-4}$ ${\rm s^{-1}}$), the eMAC wave forcing can be considered as standing waves in MITgcm simulations. At present, there is an absence of observation with which to verify the amplitude of atmospheric heating induced by eMAC wave perturbations. In this study, scaling analysis is employed. For a Jupiter-like gaseous giant, internal heat flux $\sim$ 10 W ${\rm m^{-2}}$ is expected \citep{san-2004,zaghoo,li-heat}. The vertical velocity is about 1-2 m ${\rm s^{-1}}$ and the amplitude of heating rate $10^{-7}$ K ${\rm s^{-1}}$ is generally appropriate at the bottom of weather layer \citep{showman2019}. 

Globally isotropic random perturbations $S_{\rm rand}$ are introduced  from~0.7~bar to~0.2~bar with the amplitude exponentially decreasing from~$10^{-7}$~${\rm K/s}$ to zero, suggesting a strong latent heat release of ammonia clouds in this region \citep{hueso}. $S_{\rm rand}$ is with the same form as Eqn. (3) of \citet{lian}.

The drag timescale, denoted by $\tau_{\rm drag}$, is introduced at both the bottom and the top of the atmospheric model. The drag timescale at the bottom is considered to be a free parameter with a wide adjustment range. It represents the Lorentz forces that break jets penetrating deeply into the interior of a gaseous giant. In our atmospheric simulations, where the background pressure exceeds 10 bar, the process is characterised by the relaxation of wind speed to zero over a timescale $\tau_{\rm drag}=10^7$ s. Within the top 10 grid levels, $\tau_{\rm drag}=10^5$ s denotes an extremely strong drag effect, which avoids unphysical atmospheric wave reflections caused by the grid upper boundary.

\section{Result}
\label{sec:result}
\begin{figure}[ht]
    \centering  
    \subfigure[]{  
        \includegraphics[width=0.47\textwidth]{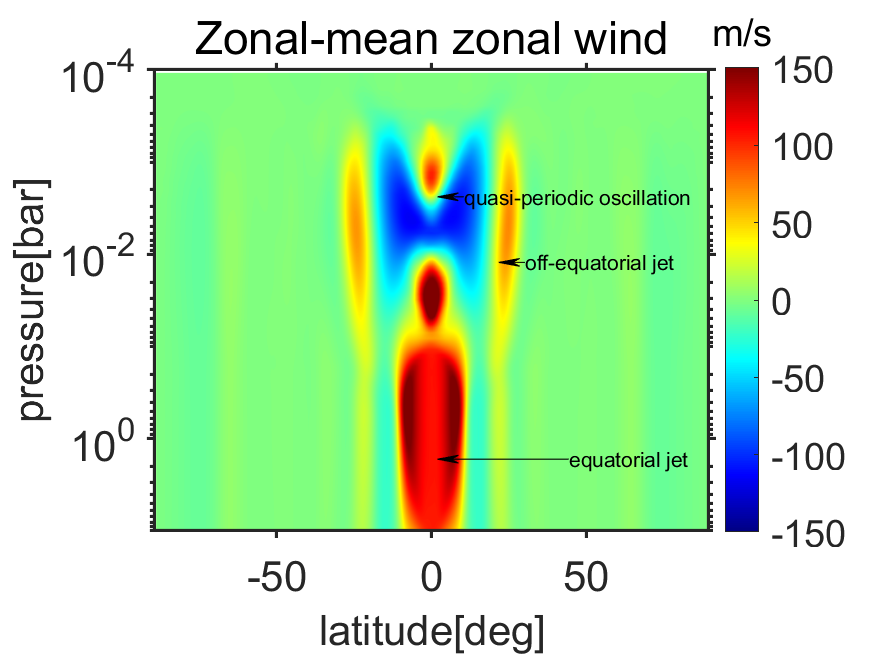}  
        \label{fig:21}  
    }  
    \hfill  
    \subfigure[]{  
        \includegraphics[width=0.47\textwidth]{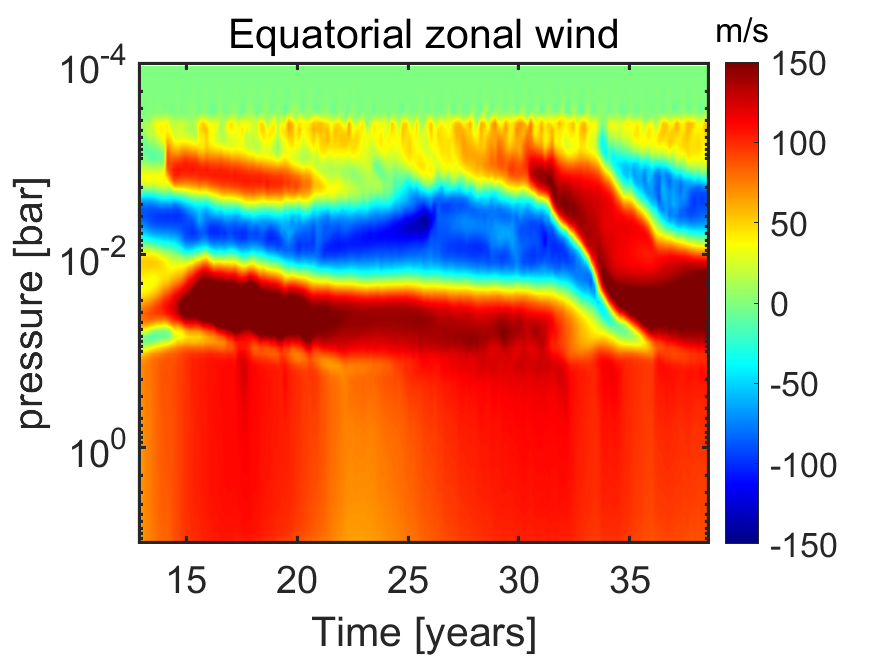}
        \label{fig:22}  
    }  
    \caption{Zonal wind structure with a stable-layer thickness of~3500~km. (a) A snapshot of zonal-mean zonal wind on the latitude-pressure diagram at year 15 of the model run. Eastward winds are represented by positive values, while westward winds are represented by negative values. (b) A time-pressure diagram at the equator, showing the equatorial jet evolution. The quasi-periodic oscillation is located in the stratosphere, with a stable eastward jet in the troposphere.}  
    \label{fig:2}  
\end{figure}

We performed simulations with varying the stable-layer thickness $H$ of 3500, 2000, 1000, and 0 km, allowing each to continue until achieving dynamic equilibrium. Figure \ref{fig:2} shows simulated zonal wind field contours with $H=3500$ km. Figure \ref{fig:21} shows zonal-mean zonal wind on a latitude-pressure diagram at year 15 of the model run. The equatorial stratosphere ($<$ 0.1 bar) and troposphere ($\ge$ 0.1 bar) exhibit distinct differences. The equatorial stratosphere displays stacked jets, indicative of quasi-periodic oscillation, whereas the troposphere demonstrates steady equatorial eastward jets. Figure \ref{fig:22} shows the evolution of equatorial jets. The equatorial stratosphere displays a quasi-periodic oscillation. The stacked jets move downward from the top of the atmosphere to the top of the troposphere, while new jets emerge at the top of the atmosphere.

\begin{figure}[ht]
    \centering   
        \includegraphics[width=0.5\textwidth]{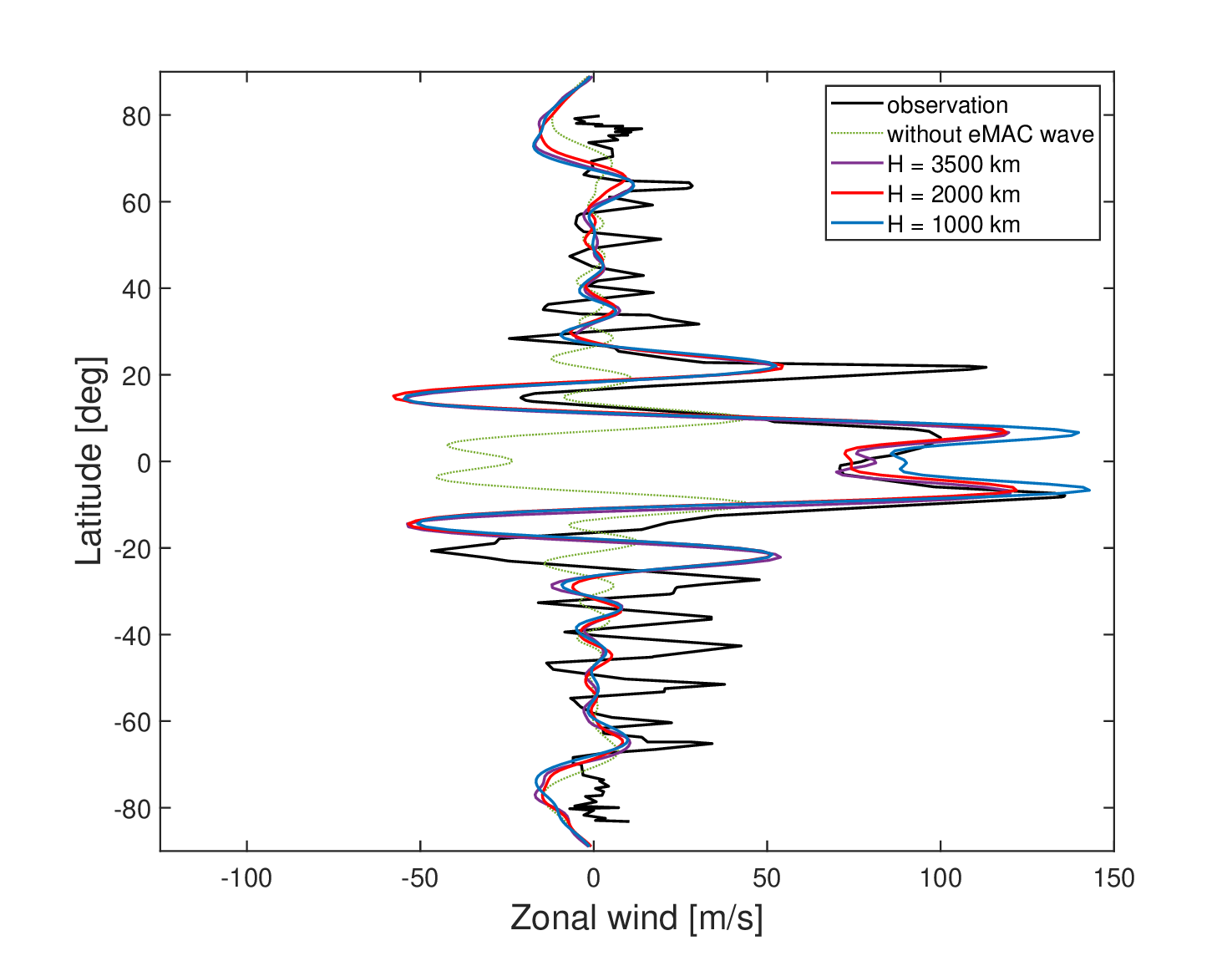}
    \caption{A sample of four generated zonal wind profiles at~0.3~bar in our simulations (colored line) with different thicknesses $H$ of stable layers. The pink dot-dashed line shows the simulation result without stable layers and eMAC wave forcing. The observed Jupiter's cloud-level wind is shown in black \citep{johnson}.}  
    \label{fig:3}  
\end{figure}

{Figure \ref{fig:3} plots the mean zonal wind field at the cloud level \citep{tollefson2017,johnson}}. The configuration of the intense equatorial eastward jet, clearly evident, closely corresponds to Jupiter's observations in low-latitude regions. {It can be noticed that within the experimental range of stable layer thicknesses (1000-3500 km), the choice of the thickness does not impact the resultant wind field significantly.} However, the eMAC wave's influence is necessary for equatorial eastward jets. {The equatorial jets show westerly without the eMAC wave's forcing (green dashed line).}

It should be noticed that Reynolds stress with eddies plays an important role in equatorward momentum transport and in generating eastward equatorial jets. {The zonal-mean zonal flow is diagnosed by the equation \citep[e.g.][]{reynolds,andrews}}:

\begin{equation}
    \frac{\partial \overline{u}}{\partial t}=-\frac{1}{a\sin^2\theta}\frac{\partial}{\partial \theta}(\overline{u'v'}\sin^2\theta)-\frac{\partial}{\partial p}(\overline{u'w'})+...
    \label{eqn:accl}
\end{equation}
{where $u, v, w$ represent zonal, meridional, and vertical velocity, primed means `perturbed' and overline means zonal-mean.} 

\begin{figure}[ht]
    \centering  
        \includegraphics[width=0.8\textwidth]{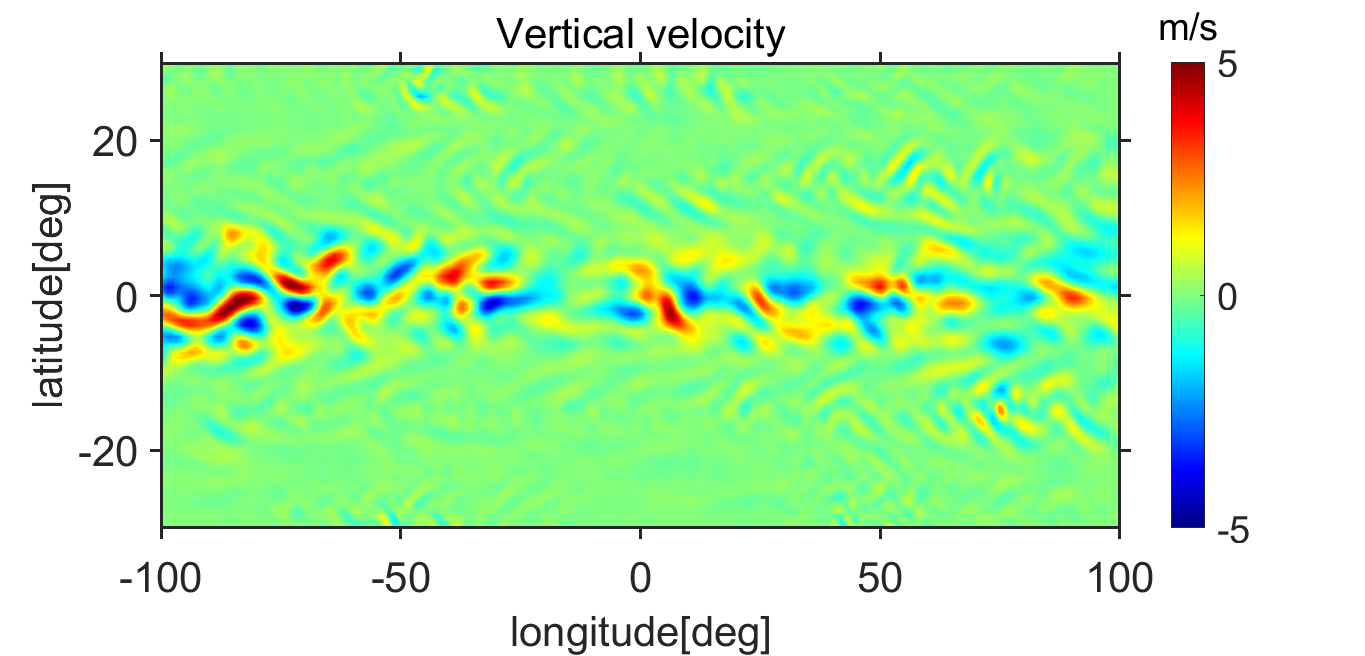}     
    \caption{Vertical velocity in the equatorial region at 1 bar. The Chevron pattern exhibits a northwest-southeast tilt in the northern hemisphere.}  
    \label{fig:4}  
\end{figure}

Figure \ref{fig:4} illustrates the vertical velocity distribution in the equatorial region at 1 bar. The vertical velocity is inherently smaller than the horizontal velocity, which enables the revelation of more nuanced wind field structures. The vertical velocity shows a chevron pattern with ``northwest-southeast'' tilt in the northern hemisphere and ``southwest-northeast'' tilt in the southern hemisphere. The chevron pattern shows evidence of an equatorial-trapped Rossby–Kelvin wave, which is a classical scenario of the Matsuno-Gill mode \citep{mat-1966,gill}. The equatorial Kelvin wave propagates eastward, while the off-equatorial Rossby waves propagate westward. Interestingly, this structure has been observed in the equatorial region of Jupiter using both ground-based telescopes and the Hubble Space Telescope. It has been observed that some deeper dark clouds exhibit an equatorially symmetric chevron structure that can persist for weeks or longer \citep{legarrera,fletcher-2018}.

The corresponding horizontal wind field of Matsuno-Gill mode shows $\overline{u'v'}<0$ for the northern hemisphere and $\overline{u'v'}>0$ for the southern hemisphere, which aligns with the first term of Eqn. \ref{eqn:accl}. This indicates that momentum transport southward in the northern hemisphere and northward in the southern hemisphere results in an eastward acceleration of the equatorial jet. The zonal inhomogeneities introduced by the perturbation of eMAC waves cause the baroclinic instabilities, and then the temperature-induced baroclinic potential energy is converted into kinetic energy of the jets via eddies \citep{s11}. 

These results show that the jets in the atmosphere of gaseous giants may have disparate origins. We find that the steady equatorial eastward jets in the troposphere originate from the convergence of momentum towards the equator due to waves' zonal inhomogeneity, which can be a consequence of the upward propagation of eMAC wave disturbances generated in the stable layers of gaseous giants' interior. In contrast, the quasi-periodic oscillation in the stratosphere is driven by atmospheric perturbations due to the latent heat release from clouds.

\section{Discussion}
\label{sec:dis}
This study, for the first time, considers the eMAC wave effects in gaseous giants' atmospheric dynamics and successfully reproduces the equatorial eastward jet. The eMAC wave perturbation provides zonal inhomogeneities needed by prograde equatorial jets. In most cases, the zonal inhomogeneities are supplied by strong external irradiation contrast of the hot Jupiter \citep{s15,plur,lesjak}, or by the dry convection with an infinitely large Prandtl number \citep{berco-2020,lange-2021}. However, these two mechanisms are not present on a Jupiter-like gaseous giant. 

{Noteworthy, in addition to the possibility of HRL forming stable stratification, the existence of other stable stratifications within Jupiter cannot be ruled out. \citet{christensen-2024} mentions a shallower stable stratification at a position of 0.97 fraction of radius, to provide an explanation for the depth decay of Jupiter's zonal jets. The conductivity at the location is significantly below 1 \citep{french-etal-2012}, making it impossible generate the eMAC waves. \citet{mil-2024} proposes a dilute core with stable stratification in Jupiter. Indeed, the eMAC waves may exist within Jupiter’s dilute core. However, in comparison with the HRL layer, the properties of the dilute core are not fully known, thereby impeding the precise estimation of the stable stratification. Our research presents an exploratory study that primarily examines the potential influence of internal structure on the atmosphere. The stable stratification in the dilute core warrants further investigation in the future.}

Our research suggests that eMAC wave solutions with $m=1$ and $n=0$ have the potential to generate the zonal inhomogeneities in the equatorial weather layer, which result in the equatorward momentum convergence to trigger the eastward jets. This is a qualitative result, as existing studies cannot provide further evidence of dynamics beneath Jupiter's cloud level. Our results show that the hydromagnetic waves within the deep interior have the potential to influence shallow atmospheric dynamics, which has never been considered before. Besides, this study helps solve the dynamical balances of Jupiter's baroclinic atmosphere. And, it should be noted that our findings are not inconsistent with the deep atmospheric model, which predicts barotropic fluid columns aligned with the rotation axis, as per the Taylor-Proudman theorem \citep{zhang-b,kaspi-etal-2009}. Both effects may have contributed simultaneously to the formation of equatorial jets.

Despite the absence of directly observed Jupiter's features——infrared anomalies of eMAC wave perturbations, some Jupiter's cloud structures may be associated with these perturbations \citep{legarrera}. Furthermore, the anomalous elongation of Jupiter's quasi-quadrennial oscillation \citep{antu} may also be associated with the eMAC wave perturbations. To our current knowledge, the eMAC-induced infrared anomalies have not been observed, which may be due to the eMAC perturbation lying beneath clouds. Further detailed infrared observations, complemented by the selection of specific atmospheric windows, have the potential to provide additional evidence of Jupiter's eMAC wave perturbations in the future.  

\begin{figure}[ht]
    \centering   
        \includegraphics[width=0.8\textwidth]{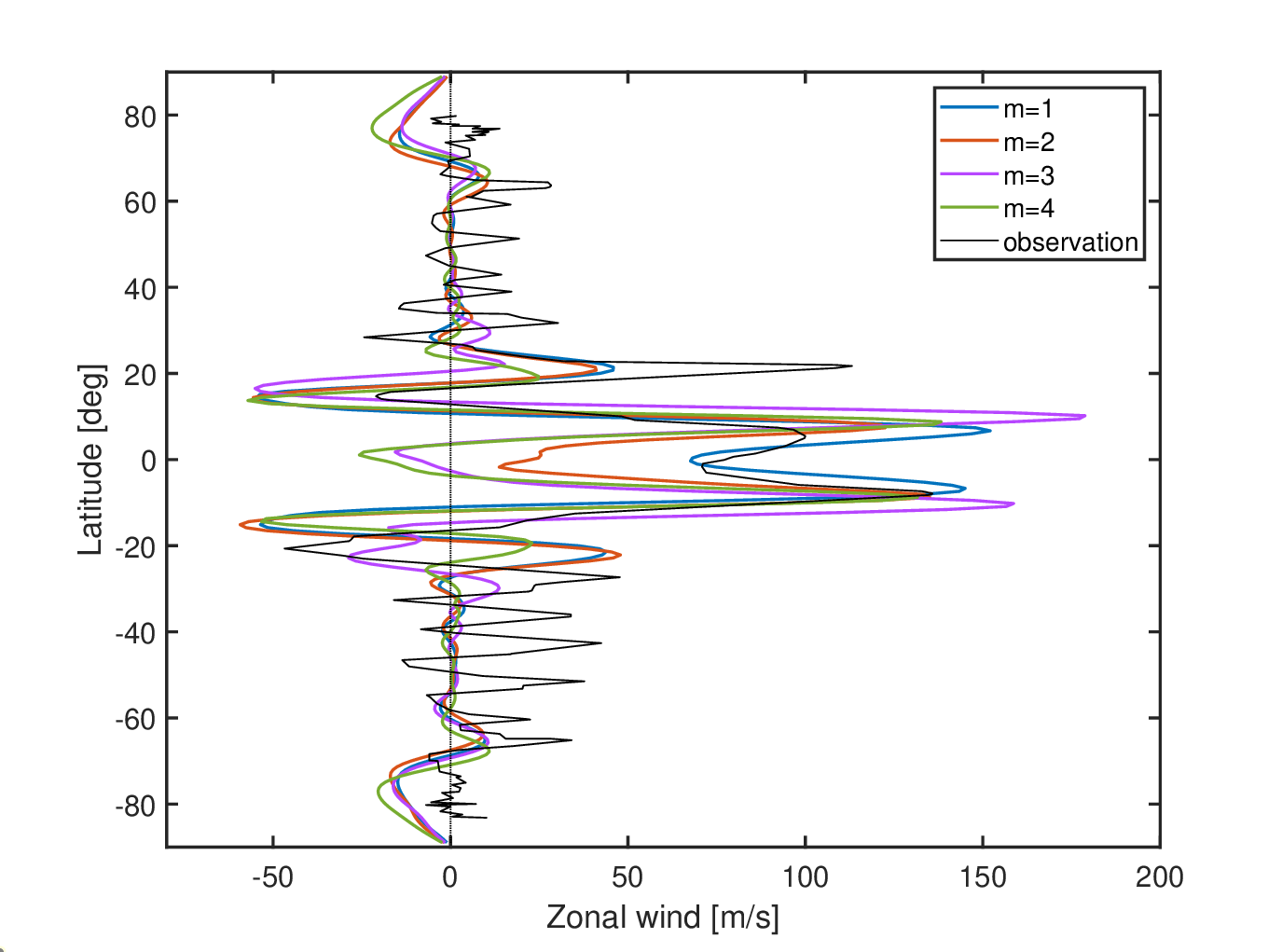}   
    \caption{A sample of four generated zonal wind profiles at~0.3~bar in our simulations (colored line) with different zonal wavenumber $m$ of eMAC waves. The thickness of the stable layer is~3500~km. The observed cloud-level wind is shown in black \citep{johnson}.}  
    \label{fig:zwno}  
\end{figure}

The ambiguity in this framework concerns the zonal wavenumber and the amplitude of eMAC waves. Indeed, we can reproduce the jets as observations corresponding to the eMAC modes with $m=1$. We also test the zonal wavenumber $m=2,3,4$ in our simulations.  Figure \ref{fig:zwno} stipulates that only perturbations with a zonal wavenumber of 1 can effectively drive strong equatorial eastward jets in the atmosphere \citep{showman2009,s11}. Existing theories cannot confine the zonal wavenumber, such as the Earth has observed wavenumber $m=7$, which has not been explained \citep{chi-2020}. We propose two mechanisms favoring small wavenumbers:  (1) Zonal heterogeneity in Jupiter's magnetic field structure, notably the low-latitude Great Blue Spot exhibiting predominant wave-1 symmetry \citep{moore2018,moore2019};  (2) Wavenumber-selective dissipation via radiative damping, where short-wavelength modes are preferentially dissipated due to their shorter thermal relax timescales \citep{roger-2005}. 

Our work is primarily a framework, and focused on mechanistic elucidation; More precise parameterization, such as the eMAC wave amplitudes, is deferred to future Jovian missions. The new constraint of HRL naturally explains the atmospheric circulation differences between gaseous and ice giants: Jupiter and Saturn have stable internal layers, manifesting equatorial eastward (prograde) jets. In contrast, Uranus and Neptune, which lack internal stable layers \citep{helled-2018-2,teanby}, {or have internal stable layers at low-conductivity regions \citep{amoros}, which cannot generate eMAC waves}, show equatorial westward (retrograde) jets. We suggest that the spatial distribution of a zonal wind field can be seen as a diagnostic of the very deep interior of giant planets, which may play an essential role in future planetary exploration.

\begin{acknowledgments}
We thank Prof. Jun Yang, Prof. Feng Ding from Peking University, and Prof. Xianyu Tan from Shanghai Jiao Tong University, for their helpful comments. Numerical simulations were conducted at the High-performance Computing Platform of Shanghai Astronomical Observatory, Chinese Academy of Sciences. This work was supported by the National Key R$\&$D Program of China (grant No. 2025YFF0512400) and the National Natural Science Foundation of China (grant Nos. 12425306, 12250013, 42405129, and 12273095).
\end{acknowledgments}

\begin{contribution}
Y.C. Lian, P.S. Duan, and D.L. Kong designed research; Y.C. Lian, P.S. Duan, and D.L. Kong performed research; Y.C. Lian ran the simulation; and Y.C. Lian wrote the paper.
\end{contribution}

\appendix

\section{Generating the eMAC waves in HRL (0.83-0.88 fraction of radius)}
\label{sec:A_emac}
In our framework, the eMAC waves are treated as small perturbations superimposed on a background state, that velocity $\textbf{u}=\textbf{u}_0+\textbf{u}'$, magnetic field $\textbf{B}=\textbf{B}_0+\textbf{B}'$, pressure $p=p_0+p'$, and density $\rho=\rho_0+\rho'$. The linearized momentum equation for magnetohydrodynamics velocity is:

\begin{equation}
    \frac{\partial \textbf{u}'}{\partial t}+2 \mathbf{\Omega}\times\textbf{u}'=-\frac{1}{\rho_0}\nabla p'+\frac{1}{\rho_0\mu}\textbf{B}_0 \cdot \nabla \textbf{B}'+\frac{\rho'}{\rho_0}\textbf{g}
    \label{eqn:1}
\end{equation}
where $\textbf{u}'$, $\textbf{B}'$, $p'$, $\textbf{g}$, $\mu$, $\Omega$ are three-dimensional perturbed velocity, perturbed magnetic field strength, perturbed pressure, gravity, permeability of free space, and angular velocity, respectively. The fluid in HRL is treated as an incompressible fluid, that $\nabla \cdot \textbf{u}'=0$.

Then, we get the linearized magnetic induction equation from the Ohmic equation and Maxwell equations:

\begin{equation}
    \frac{\partial \textbf{B}'}{\partial t}=\textbf{B}_0\cdot\nabla\textbf{u}'+\eta\nabla^2\textbf{B}'
    \label{eqn:2}
\end{equation}
where $\eta=\frac{1}{\mu\sigma}$ is the magnetic diffusivity and $\sigma$ is the electrical conductivity. 

The momentum (Eqn.~\ref{eqn:1}) and induction (Eqn.~\ref{eqn:2}) equations need to be applied in spherical coordinates ($r,\theta,\lambda$), where $r$ refers to radius, $\theta$ refers to colatitude, and $\lambda$ is longitude , with the unit vector ($\vec{\mathbf{r}},\vec{\mathbf{\theta}},\vec{\mathbf{\lambda}}$). Gravity points in the negative radial direction ($\mathbf{g}=-g_1 \vec{\mathbf{r}}$), where $g_1$ is gravity in HRL, and the rotation velocity points north along the axis of rotation. {A simpliﬁcation, which is frequently used in Earth's liquid core research \citep{buff-2019}, is possible that the perturbed magnetic field ($B_{\theta}' = B_{\lambda}' = 0$) vanishes at the top of liquid core (here is at the bottom boundary of HRL), assuming that the thickness of the skin depth ($\delta=\sqrt{\frac{2}{\mu\sigma\omega}}$) is much smaller than the thickness of HRL, where $\omega$ is the wave frequency. Besides, the fluid motion of dynamo convection is much smaller scale than the eMAC wave motion, and the isotropic convection may be cancelled by time and spatial averaging, resulting in the disregarding of velocity perturbations ($u'_{\theta} = u'_{\lambda} = 0$) at the bottom boundary.} 

To simplify, the spherical coordinate ($r,\theta,\lambda$) is converted to a new coordinate ($\hat{z}, \hat{y}, \lambda$), where $\hat{z}=r-R$, that $R$ is the maximum radius of HRL, and a meridional coordinate $\hat{y}=\cos\theta$. New magnetic field perturbations are $\hat{B}_\theta=(1-\hat{y}^2)^{1/2}B_\theta'$ and $\hat{B}_\lambda=(1-\hat{y}^2)^{-1/2}B_\lambda'$. The solution of magnetic perturbations is expressed on the latitude-longitude plane as:

\begin{equation}
    \hat{B}_{\theta,\lambda}(t,\hat{z},\hat{y},\lambda)=\tilde{B}_{\theta,\lambda}(\hat{y})\sin(k_1\hat{z})e^{{\rm i}(m\lambda-\omega t)}
    \label{eqn:3}
\end{equation}
where $m$ is the zonal wavenumber, $k_1$ is the vertical wavenumber of eMAC wave, and $\tilde{B}$ is the amplitudes of the perturbed magnetic field. 

Substitute $\hat{B}_\theta$ and $\hat{B}_\lambda$ into Eqn.~\ref{eqn:1} and \ref{eqn:2} in spherical coordinates, two govern equations shown here \citep{buff-2019}:
\begin{equation}
   (\frac{\partial^2}{\partial \hat{y}^2}-(M-I)(1-\hat{y}^2)^{-1})\tilde{B}_\theta=C\hat{y}{\rm i}\tilde{B}_\lambda+{\rm i}m\frac{\partial \tilde{B}_\lambda}{\partial \hat{y}}
   \label{eqn:4}
\end{equation}
\begin{equation}
    (-m^2-(M-I)(1-\hat{y}^2))\tilde{B}_\lambda=-C\hat{y}{\rm i}\tilde{B}_\theta+{\rm i}m\frac{\partial \tilde{B}_\theta}{\partial \hat{y}}
    \label{eqn:5}
\end{equation}

where $B_r$ is the radial magnetic field strength, $N_1=\sqrt{-\frac{g_1}{\rho_0}\frac{d\rho_0}{d\hat{z}}}$ is the buoyancy frequency of HRL, $V_a=\frac{B_r}{\sqrt{\rho_0\mu}}$ is the Alfv${\rm \acute{e}}$n wave phase speed, $C=\frac{2\Omega\omega k_1^2R^2}{N_1^2}$, $I=\frac{\omega^2k_1^2R^2}{N_1^2}$ and $M=\frac{V^2_ak_1^4R^2\omega}{N_1^2(\omega+{\rm i}\eta k_1^2)}$. Eliminating other perturbation variables, the equation shows:

\begin{equation}
(1-\hat{y}^2)\frac{\partial^2\tilde{B}_\theta}{\partial \hat{y}^2}
-2\hat{y}\frac{1-\Lambda^2+2\Lambda^2\hat{y}^2}{1+\Lambda^2\hat{y}^2}\frac{m^2}{F}\frac{\partial\tilde{B}_\theta}{\partial \hat{y}}+\left[\frac{C^2\hat{y}^2}{M}+\frac{mC}{M}-M-\frac{m^2}{1-\hat{y}^2}+\frac{2m(1-\Lambda^2+2\Lambda^2\hat{y}^2)\hat{y}^2}{1+\Lambda^2\hat{y}^2}\frac{C}{F}\right]\tilde{B}_\theta=0
\label{eqn:6}
\end{equation}
where $F=m^2+(1-\hat{y}^2)M$. The radial magnetic field can be generally approximated by $B_r^2=B_{r_{eq}}^2(1+\Lambda^2\hat{y}^2)$ and $M=M_{\rm eq}(1+\Lambda^2\hat{y}^2)$, where the subscript ${\rm eq}$ means equator.

Adopting the stable-layer thickness $H\sim$~3000~km and a Jupiter-like gaseous giant parameters in Figure \ref{fig:e1}, it is possible to estimate that $C\sim 3\times10^{-6}$, $M\sim4\times10^{-7}$ with the eMAC vertical wavenumber $k_1$, which is given by $k_1=j\pi/H$ and $j=1,2,3...$. Because $\lvert M \rvert \ll m^2$ and $\lvert M \rvert \ll m^2/(1-\hat{y}^2)$, the Eqn.~\ref{eqn:6} will be:

\begin{equation}
\begin{aligned}
  \frac{\partial^2\tilde{B}_{\theta \hat{y}}}{\partial \hat{y}^2} 
+[(\Lambda^2\hat{y}-\hat{y}-2\Lambda^2\hat{y}^3)^2-\frac{(1-\hat{y}^2)(1+\Lambda^2\hat{y}^2)(\Lambda^2-1-6\Lambda^2\hat{y}^2)}{(1-\hat{y}^2)^2(1+\Lambda^2\hat{y}^2)^2}+(\frac{C^2\hat{y}^2}{M_{\rm eq}} \\
+\frac{mC}{M_{\rm eq}})\frac{1}{(1-\hat{y}^2)(1+\Lambda^2\hat{y}^2)}-\frac{m^2}{(1-\hat{y}^2)^2}]\tilde{B}_{\theta \hat{y}}=0
\end{aligned}
\label{eqn:7}
\end{equation}

where $\tilde{B}_{\theta \hat{y}}=\sqrt{(1-\hat{y}^2)(1+\Lambda^2\hat{y}^2)}\tilde{B}_\theta$. $\hat{y}$ is small enough at low latitudes that the higher-order terms of the series expansion associated with $\hat{y}$ can be rounded off and then finally arrive at the Weber equation \citep{buff-2019}:

\begin{equation}
    \frac{\partial^2\tilde{B}_{\theta \hat{y}}}{\partial \hat{y}^2}-(\alpha_2\hat{y}^2-\alpha_0)\tilde{B}_{\theta \hat{y}}=0
    \label{eqn:8}
\end{equation}
 The solution to Eqn.~\ref{eqn:8} is \citep{duan-2023}:
\begin{equation}
\tilde{B}_{\theta \hat{y}_n}(\hat{y})=\sqrt{\frac{\tilde{\alpha}}{\sqrt{\pi}2^nn!}}e^{-1/2\tilde{\alpha}^2\hat{y}^2}H_n(\tilde{\alpha}\hat{y})
\label{eqn:9}
\end{equation}
where $\tilde{\alpha}=\alpha_2^{1/4}$ and $H_n$ is Hermite polynomial solution. For the Hermite polynomial degree $n$, $\alpha_2^{-1/2}\alpha_0=2n+1$ ($n=0,1,2, ...$) is the sufficient and necessary condition that Eqn.~\ref{eqn:8} owns the Hermite polynomial solution \citep{knezek}. Finally, $\alpha_0$ and $\alpha_2$ express as:

\begin{equation}
    \alpha_0=\frac{mC}{M_{\rm eq}}-(m^2-1)-\Lambda^2
    \label{eqn:10}
\end{equation}
\begin{equation}
    \alpha_2=-\frac{C^2}{M_{\rm eq}}-\frac{mC}{M_{\rm eq}}(1-\Lambda^2)+2(m^2-1)-2\Lambda^2(\Lambda^2+1)
    \label{eqn:11}
\end{equation}

From Eqn.~\ref{eqn:10}, Eqn.~\ref{eqn:11} and the specific solution ${\alpha_0}=(2n+1)\sqrt{\alpha_2}$, we have: 

\begin{equation}
    D_n\omega^2+E_n\omega+F_n=0
    \label{eqn:12}
\end{equation}
where $D_n=\frac{\rho_0 \mu m^2(\omega+{\rm i}\eta k_1^2)^2}{B_{r_{eq}}^2k_1^4\omega^2}+(2n+1)^2\frac{R^2(\omega+{\rm i}\eta k_1^2)}{N_1^2}$ and $E_n=\frac{m(\omega+{\rm i}\eta k_1^2)}{2\Omega k_1^2\omega}[2(1-m^2-\Lambda^2)+(2n+1)^2(1-\Lambda^2)]$ and $F_n\sim\frac{B_{r_{eq}}^2}{4\Omega^2\rho_0\mu}[(1-m^2-\Lambda^2)^2-2(2n+1)^2(m^2-1-\Lambda^4-\Lambda^2)]$. Keeping the imaginary part of Eqn.~\ref{eqn:12} equals 0, we have:

\begin{equation}
    A_n\alpha_d+\eta k_1^2G_n=0
    \label{eqn:13}
\end{equation}
where $\alpha_d=-\eta k_1^2 G_n/A_n$ is the damping rate, $A_n=[\frac{2m^2\rho_0\mu}{B_{r_{eq}}^2k_1^4}+2(2n+1)^2\frac{R^2}{N_1^2}]\omega+\frac{m}{2\Omega k_1^2}[2(1-m^2-\Lambda^2)+(2n+1)^2(1-\Lambda^2)]$ and $G_n=[\frac{2m^2\rho_0\mu}{B_{r_{eq}}^2k_1^4}+(2n+1)^2\frac{R^2}{N_1^2}]\omega+\frac{m}{2\Omega k_1^2}[2(1-m^2-\Lambda^2)+(2n+1)^2(1-\Lambda^2)]$. In a waveguide of thickness $H$ with a vertical wavenumber $k_1=j\pi/H$, $j=1,2,3... $, the damping effect is insignificant only when $j=1$ \citep{duan-2023}.

The real parts of above coefficients are ${\rm Re}(D_n)=D_{n,re}=\frac{\rho_0 \mu m^2H^4}{B_{r_{eq}}^2\pi^4}+(2n+1)^2\frac{R^2}{N_1^2}$ and ${\rm Re}(E_n)=E_{n,re}=\frac{mH^2}{2\Omega\pi^2}[2(1-m^2-\Lambda^2)+(2n+1)^2(1-\Lambda^2)]$ and ${\rm Re}(F_n)=F_{n,re}\sim\frac{B_{r_{eq}}^2}{4\Omega^2\rho_0\mu}[(1-m^2-\Lambda^2)^2-2(2n+1)^2(m^2-1-\Lambda^4-\Lambda^2)]$ with $j=1$. The solution of Eqn.~\ref{eqn:12}, which is the eigenfrequency of the eMAC wave, is given by:

\begin{equation}
\omega=\frac{-E_{n,re}\pm\sqrt{E_{n,re}^2-4D_{n,re}F_{n,re}}}{2D_{n,re}}
\label{eqn:14}
\end{equation}

\section{Wave upward propagating through MoHL (0.88-0.98 fraction of radius)}
\label{sec:A_MoHL}
Treating as small perturbations superimposed on a background state, the linearized fluid equations take on the following form:

\begin{equation}
    \frac{\partial \textbf{u}'}{\partial t}+2\mathbf{\Omega}\times\textbf{u}'=-\frac{1}{\rho_0}\nabla p'+\frac{\rho'}{\rho_0}\mathbf{g}
    \label{eqn:15}
\end{equation}
\begin{equation}
\frac{\partial \rho'}{\partial t}+\nabla\cdot(\rho_0 \mathbf{u}')=0
\label{eqn:16}
\end{equation}
where $\mathbf{u}'$, $p'$, $\rho'$, $\rho_0$, $\mathbf{g}$, $\mathbf{\Omega}$ are three-dimensional perturbed velocity, perturbed pressure, perturbed density, background density, gravity, and angular velocity, respectively. The above equations are applicable to a compressible fluid in MoHL. 

The focus is on perturbations in the equatorial region. After this, a cylindrical coordinate ($s, \lambda, l$) system is considered, where $s$ denotes radius, $\lambda$ denotes longitude, and $l$ denotes latitude. The unit vector $\vec{\mathbf{s}}$ points in the radial direction, $\vec{\mathbf{\lambda}}$ points in the longitudinal direction, and $\vec{\mathbf{l}}$ points in the latitudinal direction. Gravity points in the negative radial direction ($\mathbf{g}=-g_2\vec{\mathbf{s}}$), where $g_2$ is gravity in MoHL, and the rotation velocity parallels the latitudinal direction ($\mathbf{\Omega}=\Omega\vec{\mathbf{l}}$). The radial lower boundary is set to the surface of the high-conductivity metallic hydrogen layer with $\mathbf{u}'=0$, while the upper surface is set to the planetary surface with $p'=0$ and $\rho'=0$. Motion with time scales much longer than the rotation period is invariant in the rotation vector direction \citep{busse-1970}, so the waves show the two-dimensional nature as a lack of motion in the $l$ direction, that $\mathbf{u}'=u_s'(s,\lambda,t)\vec{\mathbf{s}}+u_{\lambda}'(r,\lambda,t)\vec{\mathbf{\lambda}}$.

The thickness of MoHL is equivalent to 10$\%$ of the radius. This layer can be regarded as a thin annular shell. Therefore, the cylindrical coordinate ($s, \lambda, l$) is converted to a new coordinate ($z, \lambda, l$), and the two-dimensional divergence ($\frac{1}{s}\frac{\partial (s u_s')}{\partial s}+\frac{1}{s}\frac{\partial u_{\lambda}'}{\partial \lambda}$) could be simplified as ($\frac{\partial u_z'}{\partial z}+\frac{1}{a}\frac{\partial u_{\lambda}'}{\partial \lambda}$), and the gradient ($\frac{\partial p'}{\partial s} \vec{\mathbf{s}}$, $\frac{1}{s}\frac{\partial p'}{\partial \lambda} \vec{\mathbf{\lambda}}$) could be simplified as ($\frac{\partial p'}{\partial z} \vec{\mathbf{s}}$, $\frac{1}{a}\frac{\partial p'}{\partial \lambda} \vec{\mathbf{\lambda}}$), where $z=s-a$, that $a$ is the equatorial radius:

\begin{align}
    \frac{\partial u_z'}{\partial t}-2\Omega u_{\lambda}'&=-\frac{1}{\rho_0} \frac{\partial p'}{\partial z}-\frac{\rho'}{\rho_0} g_2 \label{eqn:17}\\
    \frac{\partial u_{\lambda}'}{\partial t}+2\Omega u_z'&=-\frac{1}{\rho_0 a}\frac{\partial p'}{\partial \lambda} \label{eqn:18}\\
    \frac{\partial \rho'}{\partial t}+\rho_0(\frac{\partial u_z'}{\partial z}&+\frac{1}{a}\frac{\partial u_{\lambda}'}{\partial \lambda})+u_z'\frac{d\rho_0}{dz}=0 \label{eqn:19}
\end{align}

We consider hydrostatic balance ($dp_0/dz=-\rho_0 g_2$) and adiabatic motions:

\begin{equation}
     \frac{\partial \delta p}{\partial t}=\frac{\partial p'}{\partial t}-g_2\rho_0 u_z'=c_s^2(\frac{\partial \rho'}{\partial t}+u_z' \frac{d \rho_0}{dz})
     \label{eqn:20}
\end{equation}
where $c_s=\sqrt{\gamma p_0/\rho_0}$ is the sound speed, $\gamma=1.41$ is the specific heat ratio of diatomic gaseous and $\delta p$ is the Lagrangian pressure perturbation.

The wave solution has a form as $p'=\tilde{p}(z)e^{{\rm i}(m\lambda-\omega t)}$ where $m$ is the zonal wavenumber and $\omega$ is frequency. Substituting the solution and variables to combine the equations, then \citep{hin-2022}:

\begin{equation}
[\frac{d^2}{dz^2}-\frac{1}{H_{\rho}}\frac{d}{dz}+\frac{\omega^2-4\Omega^2}{c_s^2}+\frac{1}{H_{\rho}^2}\frac{dH_{\rho}}{dz}-(\frac{m}{a})^2(1-\frac{N_2^2}{\omega^2})+\frac{2\Omega m}{a\omega}(\frac{1}{H_{\rho}}-\frac{2N_2^2}{g_2})]\delta p=0
\label{eqn:21}
\end{equation}
where $H_{\rho}=-\rho_0(\frac{d\rho_0}{dz})^{-1}$ is the density scale height and $N_2=\sqrt{-\frac{g_2}{\rho_0}\frac{d\rho_0}{dz}-\frac{g_2^2}{c_s^2}}$ is the buoyancy frequency of MoHL. Separate the variables to obtain the Helmholtz equation in the radial direction with $\Phi=\delta p/\sqrt{\rho_0}$:

\begin{equation}
    \frac{d^2\Phi}{dz^2}+k_2^2\Phi=0
    \label{eqn:22}
\end{equation}
and the square of the vertical wavenumber:

\begin{equation}
k_2^2=\frac{\omega^2-\omega_{ac}^2-4\Omega^2}{c_s^2}-(\frac{m}{a})^2(1-\frac{N_2^2}{\omega^2})+\frac{2\Omega m}{a\omega}(\frac{1}{H_{\rho}}-\frac{2N_2^2}{g_2})
\label{eqn:23}
\end{equation}
where $\omega_{ac}=\frac{c_s}{2H_{\rho}}\sqrt{1-2\frac{dH_{\rho}}{dz}}$ is the acoustic cutoff frequency. It has been established that the perturbations initiated by eMAC waves are of a low frequency ($\omega \ll \Omega$). The buoyancy frequency $N_2$ equals zero in the isentropic stratification of MoHL. Therefore, Eqn.~\ref{eqn:23} has an approximation:

\begin{equation}
    k_2^2\approx -(\frac{\omega_{ac}}{c_s})^2-(\frac{m}{a})^2+\frac{2\Omega m}{a\omega H_{\rho}}
    \label{eqn:24}
\end{equation}
that $k_2^2>0$ indicates radial propagating and $k_2^2<0$ indicates radial decay. Subsequently, the frequency condition for the propagation of wave energy in the equatorial region in the radial direction is as follows:

\begin{equation}
    \omega<\omega_c\equiv\frac{2\Omega m}{[(\frac{\omega_{ac}}{c_s})^2+(\frac{m}{a})^2]aH_{\rho}}
    \label{eqn:25}
\end{equation}
where $\omega_c$ is the threshold frequency. $\omega<\omega_c$ indicates radial propagating and $\omega \ge \omega_c$ indicates radial decay.

\section{Atmospheric dynamics}
\label{sec:A_dyn}
The weather layer is located in the outermost shell of the gaseous giant's fluid envelope, with a thickness of only a few hundred kilometres, yet horizontal motion scales reaching thousands of kilometres. Consequently, the fluid within this layer in our simulations is governed by the following principles: 1. The thin-shell approximation is employed, whereby the surface gravity $g_{\rm s}$ is treated as constant. 2. Radial motion (which is locally equivalent to vertical motion) occurs on scales vastly smaller than horizontal motion, with vertical velocities being significantly lower than the horizontal velocities. Thirdly, the Boussinesq approximation is employed that the background density $\rho_0$ is treated as constant.

In the context of the aforementioned conditions, MITgcm solves the primitive equations in pressure coordinates:

\begin{equation}
    \frac{d\textbf{u}_h}{dt}+f\mathbf{k}\times\mathbf{u}=-\nabla_p\Phi_{\rm grav}-\frac{\mathbf{u}_h}{\tau_{\rm drag}}
\end{equation}

\begin{equation}
    \frac{\partial \Phi_{\rm grav}}{\partial p}=-\frac{1}{\rho_0}
\end{equation}

\begin{equation}
    \nabla_p\cdot\mathbf{u}_h+\frac{\partial w}{\partial p}=0
\end{equation}

\begin{equation}
    \frac{dT}{dt}-\frac{w}{\rho_0c_p}=-\frac{T-T_{\rm eq}}{\tau_{\rm rad}}+S_{\rm eMAC}+S_{\rm rand}
\end{equation}
where $\textbf{u}_h$ represents the horizontal velocity, $w$ is the vertical velocity at pressure coordinates, $c_p$ is the specific heat at constant pressure, $f=2\Omega\cos\theta$ is the Coriolis parameter, $\nabla_p$ is the horizontal pressure gradient, $\Phi_{\rm grav}$ is gravitational potential, and $\textbf{k}$ is the vertical vector. $\tau_{\rm drag}$ and $\tau_{\rm rad}$ are the drag and radiative timescale, respectively. $T_{\rm eq}$ is defined as the equilibrium temperature. $S_{\rm eMAC}$ represents the heating/cooling caused by the upward perturbation from MoHL, and $S_{\rm rand}$ is the random forcing with the same form as Eqn. (3) of \citet{lian}.

\bibliography{sample701}{}
\bibliographystyle{aasjournal}

%% This command is needed to show the entire author+affiliation list when
%% the collaboration and author truncation commands are used.  It has to
%% go at the end of the manuscript.
%\allauthors

%% Include this line if you are using the \added, \replaced, \deleted
%% commands to see a summary list of all changes at the end of the article.
%\listofchanges

\end{document}